\documentclass[12pt]{article}
\usepackage{graphicx}
\date{}

\title{\bf \small Electronic transmission of a nanowire
partly irradiated under terahertz electromagnetic field}

\author{\small Guang Hui Zhou$^{1,2}$\footnote{Corresponding
author, e-mail: ghzhou@hunnu.edu.cn} and Mou Yang$^1$}

\begin{document}
\renewcommand{\baselinestretch}{1.5}
\maketitle
\begin{center}
{\small $^1$Department of Physics, Hu'nan Normal University,
Changsha 410081, Hunan, P. R. China}

{\small $^2$International Center for Materials Physics,\\
             Chinese Academy of Science, Shenyang 110015, P. R. China}

\vspace{0.8cm}
{\bf Abstract}
\end{center}

\small We theoretically study the electronic transport of a
nanowire partly irradiated under an external terahertz (THz)
electromagnetic field. Although the electrons in the ballistic
nanowires only suffer lateral collision with photons the
reflection of electrons also takes place in this partly irradiated
case. Using free-electron model and scattering matrix approach we
showed that at resonance there exists a step decrement of $50$
percent for the transmission probability as the amplitude of field
increases to a certain volume. And the coherent structure of
transmission for the system apparently appears when the field
irradiate the middle part of nanowire only. This sensitive
transmission property of the system may be used in the THz
detection.

\vspace{0.8cm}


\vspace{0.9cm}
\baselineskip 20pt

\newpage
Since 1980's mesoscopic physics has been extensively studied due
to its potential practice in the future. Nowadays increasingly
progressive arts and crafts of nano-technologies allow men to
realize some real mesoscopic system in laboratory. One can confine
two dimensional electron gas in an ultra narrow channel on the
heterojunction of semiconductors by applying a confinement
electronic potential (two-dimensional nanowire$^1$). Recently,
three-dimensional metallic nanowire have been constructed when
probe pin apart from a metal film.$^2$ The sizes of these
low-dimensional systems are of the order of a few nanometers which
is comparable to the Fermi wave length. In ballistic regime and at
low temperatures quantum coherent effect will dominate the
electronic transport properties of these nanowires. The most basic
and important feature is that the conductance shows histogram
structure when the lateral size of nanowire varies and each step
has a height of $2e^2/h$ or integer times of it.$^3$

The electronic transport of nanowires can be affected by many
factors. The lateral shape varying in space leads to a mixture of
different modes in nanowires. If the shape varies slow enough to
satisfy adiabatic approximation, the step-like structure of
conductance is still preserved but it is totally determined by the
narrowest neck part of the whole nanowire.$^4$ Also, when a
nanowire is exposed to a static magnetic field the transverse
energy levels will be modified so the position of the conductance
steps will be altered.$^5$ Moreover, when a nanowire is irradiated
under a proper external electromagnetic field many new features
arise because of the nonelastic scattering by photons.$^{6,7}$
When the Fermi level is below the lowest transverse level in the
neck part of nanowire electrons can not go though without external
field. But under the external field irradiation electrons can
absorb energy of photons and go though this geometric
barrier.$^{6,7}$ Therefore, in the regime of barrier the electron
reflection may be induced by the combination effect of the field
and the lateral shape variation.$^{6-9}$ However, the pure
external field effect on the transport (photonconductance) has not
drawn attention yet, and this effect is important for both basic
physics knowledge and  nanotechnology application.

The typical value of transverse level separation is of the order
of 1$\sim$10 meV for both types of the nanowires.$^{6-9}$ This
corresponds to match frequencies of the order of 1$\sim$10 THz,
which are available as the development of the ultrafast laser
technology.$^{10}$ Recently there is great interest in new solid
state devices that could detect and emit THz radiation in a
selective and tunable way.$^{11-12}$ The nanowire system may be
used to detect THz radiation in such a way.

In this letter we focus our eyesight on external field induced
electron reflection by solving the problem of a straight
cylindrical nanowire (either two-dimensional semiconductor
nanowire or three-dimensional metallic one) partly irradiated
under an external THz field. When the external field transversely
irradiate a part of a nanowire the displacement symmetry along the
wire is violated so that the longitudinal momentum is not a
conservative quantity and reflection must arise in general.
However, if the full nanowire is transversely irradiated there
will be no electron reflection in the ideal case. Using the method
of time-dependent mode matching we demonstrated an interesting
field induced step-structure of electronic transmission. And this
new effect has not been reported previously for such a quantum
wire system.

The system configuration is depicted as in Fig.1. A
three-dimensional model system consists of a nanowire connecting
two reservoirs. The longitudinal z-axis is along the wire, and the
x- and y-axes are in the transverse directions. We assume the
boundary of the wire is perfect and there is no impurities in the
body, then the free electron model is adapted. The THz field
transversely irradiate the interval of $0\leq z\leq l$ of wire in
an unspecified way. The field is described by the vector potential
$\vec A(t)=(\varepsilon/\omega)\cos\omega t\cdot\hat e_y$, where
$\omega$ and $\varepsilon$ are the angular frequency and amplitude
of field, respectively, $\hat e_y$ is the unit vector in the y
direction (polarized direction). Therefore, the time-dependent
Schr\"{o}dinger equation of the field irradiated part of wire is
\begin{equation}
\left[\frac1{2m^*}(\hat p+e\vec
A)^2+U(x,y)\right]\Psi(x,y,z,t)=i\hbar\frac{\partial}{\partial
t}\Psi(x,y,z,t)\;,
\end{equation}
where $\hat p=-i\hbar\nabla$ is the momentum, $m^*$ is the
effective mass of an electron and the potential $U(x,y)$ is in the
form of either hard-wall or parabolic which confines the electron
to the wire and to the reservoirs. We assume the electronic wave
function has a separation form
\begin{equation}
\Psi_k=e^{ikz}\sum_na_n(t)\phi_n(x,y),
\end{equation}
where $k$ is the longitudinal momentum, $a_n(t)$ is the
time-dependent state amplitude and $\phi_n(x,y)$ are
eigenfunctions of the transverse motion without field, which
depend on the specified conforming potential $U(x,y)$. For
simplicity we consider only the two lowest transverse energy
levels with Fermi level between them and ignore other higher
levels in the system. Then inserting ansatz (2) into Eq.(1) we
obtain an equation of the matrix form under the weak field
approximation ($A^2\sim0$)
\begin{equation}
i\frac{\partial}{{\partial t}}\left[{\begin{array}{*{20}c}
a_1\\
a_2\\
\end{array}}\right]=k^2\left[{\begin{array}{*{20}c}
a_1\\
a_2\\
\end{array}}\right]+\left[{\begin{array}{*{20}c}
{E_1 } & {M\cos \omega t}\\
{M\cos \omega t} & {E_2 }\\
\end{array}} \right]\left[ {\begin{array}{*{20}c}
a_1\\
a_2\\
\end{array}} \right],
\end{equation}
where $\hbar=2m^*=1$ is adapted, $E_1$ and $E_2$ are eigenvalues
of the the transverse modes and the interaction energy between the
two modes $M=|\frac {2e\varepsilon}{\omega}\int
\phi_1^*\frac{\partial}{\partial y}\phi_2dxdy|\sim\varepsilon$.
Using the rotating wave approximation we finally obtain the
electronic wavefunction
\begin{equation}
\Psi(x,y,z,t)=\sum_k\left[C_{k,+}\left|+\right\rangle
e^{-i(\lambda_+ +k^2)t}+C_{k,-}\left|-\right\rangle
e^{-i(\lambda_-+k^2)t}\right]e^{ikz},
\end{equation}
where $C_{k,\pm}$ are arbitrary constants, $\lambda_{\pm}
=\pm\frac12\sqrt{\delta^2+ M^2}$, $\delta=\omega-(E_2-E_1)$ is the
detune and $|\pm\rangle$ are two eigenvectors
\begin{equation}
\left|\pm\right\rangle=e^{i(\frac{\delta}2-E_1)t}\phi_1+
A_{\pm}e^{-i(\frac{\delta}2+E_2)t}\phi_2, \qquad
A_{\pm}=\frac1M(-\delta \pm \sqrt{\delta^2+M^2}).
\end{equation}

Firstly, we consider the case of field irradiating the right part
of nanowire. The electron wavefunction in the unirradiated part
(the left part) can be written as
\begin{equation}
\Psi(x,y,z,t)=[e^{i(k_1z-Et)}+C_1e^{-i(k_1z+Et)}]\phi_1
+C_2e^{-i[k_2 z+(E+\omega)t]}\phi_2,
\end{equation}
here we assumed the electron with unity amplitude of total energy
$E$ and momentum $k_1$ emission from left, $k_1=\sqrt{E-E_1}$,
$k_2=\sqrt{E-E_2+\omega}$ and $C_1, C_2$ are constants. Now we
match the time-dependent wave functions at the boundary of
irradiated and unirradiated part to calculate the transmission of
the system (i.e., the two wave functions and their differentials
continuum at point $z=0$), and the conductance connects the
transmission probability through Landauer-B\"{u}ttiker
formula.$^{13}$ The numerically calculated transmission
probability versus with detune $\delta (\sim\omega)$ is shown in
Fig.2. In the regime of $\left|\delta\right|>0.5$, the
transmission probability remains unchanged as the situation
without external field, but it decreases rapidly from near $1$ to
$0.5$ as $\delta$ is near to zero (i.e., $\omega\simeq E_2-E_1$).
This apparent resonance peak implies that the quantum wire system
is sensitive to the frequency of external field and may be
applicable to the detection of THz irradiation.

Since the main physics may appear at the vicinity of the resonant
frequency so we only give the result of resonant case $\delta=0$
in the following sections. In this case Eq.(4) only sum over $k_+$
and $k_-$. After a straightforward manipulation the transmission
probability for the system can be evaluated analytically
\begin{equation}
T=2\left[\left|{\frac{{\sqrt{k_1k_+}}}{{k_1+k_+}}}
\right|^2\theta(k_+^2)+\left|{\frac{{\sqrt{k_1k_-}
}}{{k_1+k_-}}}\right|^2\theta(k_-^2)\right],
\end{equation}
where $k_{\pm}=\sqrt{E-E_1\mp M}$ and $\theta(x)$ is the step
function. Fig.3 shows the transmission probability vs interaction
energy $M (\sim\varepsilon)$ in the resonant case. One can see
that the transmission decreases gradually starting from $1$ when
$M$ is small, then it drops to $0.5$ rapidly as $M$ reaches to
$0.2$. After the step-like structure the transmission decreases
linearly and slowly to $0.42$ as $M$ increases. This phenomenon
can be explained as follows. When electron penetrate though the
boundary, due to the field effect the transverse modes of the
nanowire are dressed and one electron mode is split into two
time-dependent modes $|+\rangle e^{-i(M/2+k_+^2)t}$ and $|-\rangle
e^{-i(-M/2+k_-^2)t}$ with longitudinal momentums $k_+$ and $k_-$
respectively. When $M<0.2$ both $k_+$ and $k_-$ are real and
contribute to the transmission. When $M>0.2$, $k_+$ is unreal so
the mode $|+\rangle e^{-i(M/2+k_+^2)t}$ corresponds to a
evanescent (nonpropagating) channel which contributes nothing to
the transmission. Therefore the step occur at the point of $M=0.2$
when total electron energy $E=1.2$. Because electron at each mode
has two momentum $k_+$ and $k_-$, the Rabi oscillation in space
with wavelength $2\pi/(k_- -k_+)$ arises in the field part of
nanowire, and this phenomenon also has been predicated in
Ref.[7,8].

Next, we consider the case of field irradiating the middle part of
nanowire. In this case the system contains two boundaries of
irradiated and unirradiated. To obtain the total transmission
through the two boundaries with the length of field irradiated
part $l$ (see Fig.1), we use the approach recently developed by
Torres and Sa\'{e}nz.$^{14}$ This approach need to derive the
total scattering matrix which can be expressed in the transmission
matrix and reflection matrix on each boundary. Because of the
similarity of the two boundaries one has not to match the wave
function at the right boundary. The total transmission matrix is
just the anti-diagonal submatrix for the symmetry case. The detail
knowledge about this is refereed to Ref.[14] and is not presented
here. After some manipulation one can get the total transmission
matrix
\begin{equation}
t_{total}=t(1-Ur'Ur')^{-1}Ut',\;\quad \quad \;\;\;U =\left[
{\begin{array}{*{20}c}
{e^{ik_+z}}&0\\
0&{e^{ik_-z}}\\
\end{array}}\right],
\end{equation}
where $t'(t)$ is the transmission matrix from left(right) to
right(left) of the first boundary, $r'$ is the reflection matrix
from right to left of the first boundary and $U$ is propagating
matrix.$^{12}$ And the total transmission probability reads
\begin{eqnarray}
T_{total}&=&\textrm{Tr}[t^{\dag}_{total} t_{total}]\nonumber\\
&&=2\left|\frac{2k_1k_+}{e^{-ik_+l}(k_1+k_+)^2-e^{ik_+l}(k_1-
k_+)^2}\right|^2\nonumber\\
&&+2\left|\frac{2k_1k_-}{e^{-ik_-l}(k_1+k_-)^2-e^{ik_-l}
(k_1-k_-)^2}\right|^2.
\end{eqnarray}
In Fig.4 we show the calculated total transmission probability of
Eq.(9) as a function of the interaction energy $M$. One can see
that for the same reason as in Fig.3 the curve also has a step
structure at point $M=0.2$. However, it shows apparently coherence
pattern. When an electron propagate in the middle part of
nanowire, the electron occupies two time-dependent states with
longitudinal momentum $k_+$ and $k_-$ respectively. When $M>0.2$
the mode corresponding to $k_+$ become a evanescent one so that
the total transmission probability is suppressed to around 0.4. As
the length of the field irradiated middle part $l$ is an integer
time of $\pi/k_-$ the coherent peaks appear. However, when $M<0.2$
the both two modes are propagating and the total transmission is
the coherence superposition of these two modes. One can find that
the curve in Fig.4 is just the one in Fig.3 superimposed with some
coherence structure due to the interference of the forward-going
wave and backward-going wave in the middle part.

In summary, we have theoretically studied the electronic transport
property of a nanowire (or sometimes quantum wire, electron
waveguide, etc.) partially irradiated by an external THz field. We
find that the reflection will arise when the electrons suffer only
lateral collision with photons. Whether the field irradiate the
right part or the middle part of nanowire the transmission always
show an interesting step structure in the resonant case when the
amplitude of external field increases to a critical volume. But in
the later case the transmission curve apparently shows coherent
structure. This interesting electronic transport property of the
nanowire system may be useful for the detection of THz
irradiation.

This work was partially supported by the Natural Science
Foundation of Hunan under Grant No.0210138 and by the Research
Foundation of Hunan Normal University under Grant No.00206.

\newpage

\begin{figure}
\center
\includegraphics[width=3in]{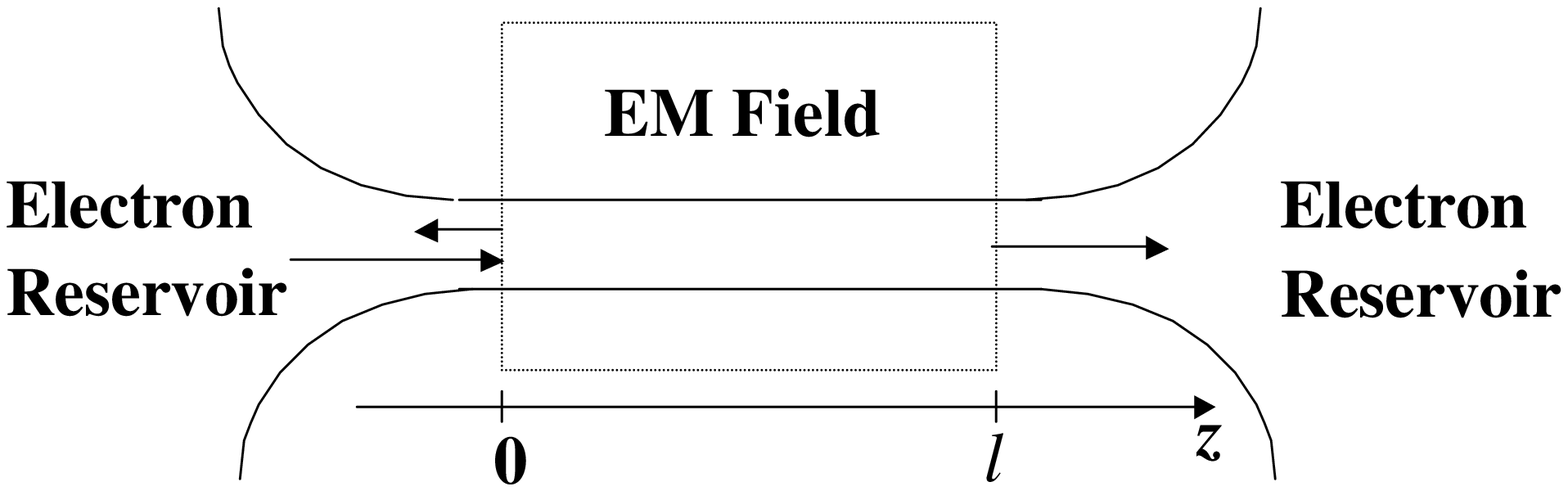}
\caption{The sketch configuration of the system when the external
field transversely irradiate the middle part of the nanowire. The
arrows denote the propagation directions of electrons.}
\end{figure}

\begin{figure}
\center
\includegraphics[width=3in]{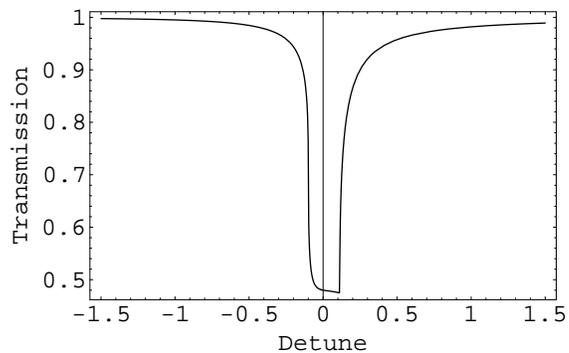}
\caption{Transmission probability vs detune $\delta$ in the case
of field irradiating the right part of the nanowire. The
parameters are specified as $E=1.2$, $ E_1=1$ and $M=0.2$.}
\end{figure}

\begin{figure}
\center
\includegraphics[width=3in]{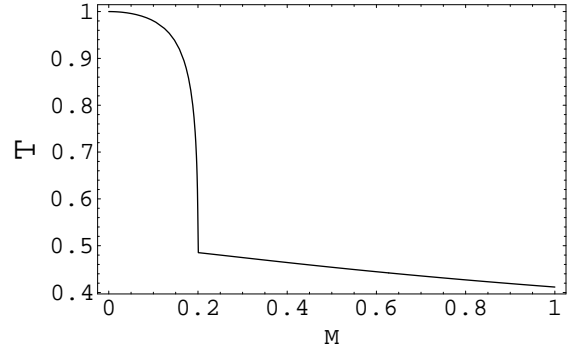}
\caption{Transmission probability vs interaction energy $M$ for
the case of field irradiating the right part of the nanowire (with
single boundariy). Parameters are specified as $E=1.2$ and
$E_1=1$.}
\end{figure}

\begin{figure}
\center
\includegraphics[width=3in]{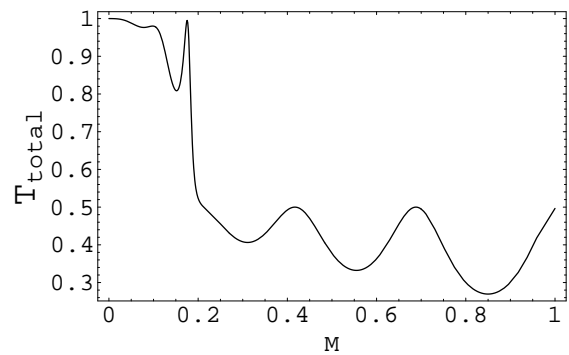}
\caption{Transmission probability vs interaction energy $M$ for
the case of field irradiating the middle part of the nanowire
(with two boundaries). $l=33$ and other parameters are the same as
in Fig.3.}
\end{figure}

\end{document}